\documentclass[11pt]{article}
\usepackage{fancyhdr}

\usepackage{amssymb}
\usepackage{setspace}
\usepackage{graphics,epsfig,cite,subfigure,amsmath,amsthm}
\usepackage[all]{xy}
\newtheorem{theorem}{Method}[section]   

\theoremstyle{definition}
\newtheorem{definition}[theorem]{Definition}   

\theoremstyle{remark}
\newtheorem{example}[theorem]{Example}        

\begin{document}

\title{Molecular Network Control Through Boolean Canalization}

\author{David Murrugarra$^a$ \and Elena S Dimitrova$^b$}
\date{}
\maketitle
\thispagestyle{fancy} 
{\footnotesize
  \centerline{$^a$Department of Mathematics,
  University of Kentucky, Lexington, KY 40506-0027 USA.}
}
{\footnotesize
  \centerline{$^b$Department of Mathematical Sciences,
Clemson, SC 29634-0975 USA}
}
%
\begin{abstract} 
Boolean networks are an important class of computational models for molecular interaction networks. Boolean canalization, a type of hierarchical clustering of the inputs of a Boolean function, has been extensively studied in the context of network modeling where each layer of canalization adds a degree of stability in the dynamics of the network. Recently, dynamic network control approaches have been used for the design of new therapeutic interventions and for other applications such as stem cell reprogramming. This work studies the role of canalization in the control of Boolean molecular networks.  
It provides a method for identifying the potential edges to control in the wiring diagram of a network for avoiding undesirable state transitions.
The method is based on identifying appropriate input-output combinations on undesirable transitions that can be modified using the edges in the wiring diagram of the network.
Moreover, a method for estimating the number of changed transitions in the state space of the system as a result of an edge deletion in the wiring diagram is presented.
The control methods of this paper were applied to a mutated cell-cycle model and to a \textit{p53-mdm2} model to identify potential control targets.
\end{abstract}
\section{Background}
A Gene Regulatory Network (GRN) is a representation of the intricate relationships among genes, proteins, and other substances that are responsible for the expression levels of mRNA and proteins.
The amount of these gene products and their temporal patterns characterize specific cell states or phenotypes. Thus GRNs play a key role in the understanding of the various functions of cells and
cellular components, and ultimately might help to design intervention strategies for the control of biological systems.
Recently, practical applications in cancer systems biology such as the identification of new
therapeutic targets has stimulated the development of computational tools that can help to identify new intervention targets.
Experimentally, the interventions are realized by manipulating the wiring diagram
of a system with the use of drugs or by gene knockouts to impact the dynamics of the system
so that it is directed towards a desired state~\cite{Choi2012,Wang:2013aa,Lee:2012aa,Wang:2014qy,Erler:2012zr}.
From the modeling perspective, the identification of intervention targets amounts to finding a set of relevant nodes and edges that can be used for performing interventions \emph{in silico}. 

Many dynamic systems theory approaches have been used over the last decades to develop computational tools for analyzing the dynamics of GRNs. As a result, a large variety of models exists today.
Boolean networks is a class of computational models in which genes can only be in one of two states: ON or OFF. 
BNs, and more general discrete models in which genes can take on more than two states, have been effectively used to model biological systems such as the yeast cell cycle network \cite{Li:2004aa}, the Th regulatory network~\cite{Mendoza:2006aa}, the \textit{lac} operon~\cite{DBLP:journals/jcb/Veliz-CubaS11}, the \textit{p53-mdm2} complex~\cite{Choi2012,Murrugarra2012,Abou-Jaoude:2009aa}, A. \textit{thaliana}~\cite{Balleza:2008aa}, and for many other systems~\cite{Davidich:2008aa, Albert:2003aa, Saadatpour:2011aa, Zhang:2008aa, Helikar:2008aa, Helikar:2013aa}.

Boolean networks (BNs) as models for GRNs were introduced by S.~Kauffmann~\cite{Kauffman:1969aa} and R.~Thomas~\cite{Thomas:1990aa}.
BNs have been proposed as a framework that does not rely on kinetic constants and therefore requires fewer parameters to estimate, which simplifies analysis. 
Boolean canalizing rules were introduced by S.~Kauffman and collaborators~\cite{Kauffman2003} and reflect the concept of canalization in evolutionary biology that Waddington pioneered in 1942~\cite{waddington1942}.
Boolean canalization has been intensively studied from the network dynamic perspective~\cite{Kauffman:2004aa,Jarrah:2007aa,Murrugarra:2012aa,Li:2013aa}.
It has been shown that networks that use only nested canalizing rules exhibit more stable dynamics compare to network using random rules~\cite{Kauffman:2004aa,Murrugarra:2011aa}.
Furthermore, it has shown that each additional layer of canalization provides a degree of stability~\cite{Layne:2012aa,Li:2013aa}. 
The Boolean functions in published models tend to have many canalizing variables~\cite{Murrugarra:2011aa,Kauffman2003}.
  
\subsection*{Boolean Networks}
A Boolean network can be defined as a dynamical system that is discrete in time as well as 
in variable states. More formally,
consider a collection $x_1, \ldots , x_n$ of
variables, each of which can take on values in the binary set $\{0,1\}$.
A Boolean network in the variables $x_1, \ldots , x_n$ is a function 
\begin{displaymath}
\mathbf{ F} = (f_1,\dots,f_n):\{0,1\}^n\rightarrow \{0,1\}^n
\end{displaymath}   
where each coordinate function $f_i$ is a Boolean function on a subset of $\{x_1,\dots,x_n\}$ which represents how the future value of the $i$-th variable depends on the present values of the variables.


The dynamical properties of a Boolean network are given by the difference equation $x(t+1)=\mathbf{ F}(x(t))$; that is, the dynamics is generated by iteration of $\mathbf{ F}$. More precisely,
the dynamics of $\mathbf{ F}$ is given by the state space graph $S$, defined as the graph with vertices in
$\mathbb{K}^n=\{0,1\}^n$ which has an edge from $x\in \{0,1\}^n$ to $y\in \{0,1\}^n$ if and only 
if $y = \mathbf{ F}(x)$. In this context, the problem of finding the states
$x\in \{0,1\}^n$ where the system will get stabilized is of particular importance. These special points of the state space are called attractors of a Boolean network and these may include steady states (fixed points), where $\mathbf{ F}(x) = x$, and cycles, where $\mathbf{ F}^r(x) = x$ for some integer number $r>1$. Attractors in Boolean network modeling might represent cell types~\cite{Kauffman:1969aa} or cellular states such as apoptosis, proliferation, or cell senescence~\cite{Huang:1999aa,DBLP:books/daglib/0024105}.
\subsection*{Canalizing functions}
A Boolean function $f(x_1,\dots,x_n):\{0,1\}^n\rightarrow \{0,1\}$ is canalizing in the variable $x_i$ with canalizing
input value $a$ and canalizing output value $b$ if $f(x_1,\dots,x_i=a,\dots,x_n) = b$. That is, once $x_i$ gets its
canalizing input, it by itself determines the output of the function regardless of the value of the other variables. The variable $x_i$ is called a canalizing variable.
\subsection*{Nested canalizing functions}
Let $\sigma$ be a permutation on the set $\{1,2,\dots,n\}$.
The function $f(x_1,\dots,x_n):\{0,1\}^n\rightarrow \{0,1\}$ is a {\it nested canalizing function} (NCF) in the 
 variable order $x_{\sigma(1)},\dots,x_{\sigma(n)}$ with {\it canalizing input values}
 $a_1,\dots,a_n\in \{0,1\}$ and {\it canalizing output values} $b_1,\dots,b_n\in \{0,1\}$ if it can be represented in the form 
\begin{displaymath}
f=\left\{
\begin{array}{l}
b_1\text{ if}\ x_{\sigma(1)} = a_1\\
b_2\text{ if}\ x_{\sigma(1)}\neq a_1,x_{\sigma(2)}= a_2\\
b_3\text{ if}\ x_{\sigma(1)}\neq a_1,x_{\sigma(2)}\neq a_2,x_{\sigma(3)}= a_3\\
\vdots\\
b_n\text{ if}\ x_{\sigma(1)}\neq a_1,\dots,x_{\sigma(n)}= a_n\\
b_{n}+1\text{ if}\ x_{\sigma(1)}\neq a_1,\dots,x_{\sigma(n)}\neq a_n.
\end{array}\right.
\end{displaymath}
Let $\alpha = \{a_1,\dots,a_n\}$ and $\beta = \{b_1,\dots,b_n\}$. Then
$f$ is said to be $\{\sigma,\alpha,\beta\}$ NCF if it is NCF in the variable order $x_{\sigma(1)},\dots,x_{\sigma(n)}$ with canalizing input values
 $\alpha$ and canalizing output values $\beta$.
\subsection*{Partially nested canalizing functions}
Any Boolean function $f(x_1,\dots,x_n):\{0,1\}^n\rightarrow \{0,1\}$ can be represented in the form: 
\begin{displaymath}
f=\left\{
\begin{array}{l}
b_1\text{ if}\ x_{\sigma(1)} = a_1\\
b_2\text{ if}\ x_{\sigma(1)}\neq a_1,x_{\sigma(2)}= a_2\\
\vdots\\
b_d\text{ if}\ x_{\sigma(1)}\neq a_1,\dots,x_{\sigma(d)}= a_d\\
g\text{ if}\ x_{\sigma(1)}\neq a_1,\dots,x_{\sigma(d)}\neq a_d
\end{array}\right.
\end{displaymath}
where either $n=d$, where $x_d$ is a terminal canalizing variable and $g$ is a constant, or
$n<d$ where $g(x_{\sigma(d+1)},\dots,x_{\sigma(n)})$ is a non-constant function and
none of the variables $x_{\sigma(d+1)},\dots,x_{\sigma(n)}$ are canalizing for $g$.
The integer $d$ is called the nested canalizing depth of $f$. Such Boolean functions are called partially nested canalizing functions (PNCFs), see~\cite{Layne:2012aa} for more details.  
\subsection*{Layers of Canalization}
A Boolean function can be represented in different forms as a nested canalizing function.
A unique representation of the function is obtained by grouping the variables in layers of canalization~\cite{Li:2013aa}.
Every Boolean function can be uniquely written as
\begin{equation}
\label{Eq:layers}
f(x_1,\dots,x_n) = M_1(M_2(\dots(M_{r-1}(M_rP_c+1)+1)\dots)+1)+b,
\end{equation}
where $M_i = \prod_{i=1}^{k_i}(x_{i_j}+a_{i_j})$, $P_c$ is a polynomial
with no canalizing variables, and $k = k_1+\cdots+ k_r$ is the canalizing depth.
Each variable $x_i$ appears in exactly one of the $M_1,M_2,\dots,M_r,P_c$.
The proof of this property is given in~\cite{He:2015aa}. The number $r$ in 
Equation~\ref{Eq:layers} is called the \emph{layer number} of $f$.
\begin{example}\label{ex:layers}
Consider the Boolean functions $f_1$, $f_2$, and $f_3$ with truth tables given at Table~\ref{table:truthtable}.
The layers representation for $f_1$ is
$$f_1(x_1,x_2,x_3) = x_1x_3(x_2+1)+1,\text{ where }M_1 = x_1x_3(x_2+1),$$ 
for $f_2$ is
$$f_2(x_1,x_2,x_3) = (x_1+1)[x_2(x_3+1)+1]+1,\text{ where }M_1=x_1+1,\ M_2=x_2(x_3+1),$$
and for $f_3$ is
$$f_3(x_1,x_2,x_3) = (x_1+1)(x_2+x_3+1),\text{ where }M_1=x_1+1,\ P_c=x_2+x_3+1.$$
Thus $f_1$ has layer number equal to 1, $f_2$ has layer number equal to 2, and $f_3$
has layer number 1. The polynomial $P_c$ does not have canalizing variables. 
\end{example}
\begin{table}
  \centering
  \begin{tabular}{|c | c|}
\hline
$x_1\ x_2\ x_3$&$f_1$\\\hline
0 0 0& 1\\
0 0 1& 1\\
0 1 0& 1\\
0 1 1& 1\\
1 0 0& 1\\
1 0 1& 0\\ 
1 1 0& 1\\
1 1 1& 1\\
\hline
\end{tabular} 
\begin{tabular}{|c | c|}
\hline
$x_1\ x_2\ x_3$&$f_2$\\\hline
0 0 0& 0\\
0 0 1& 0\\
0 1 0& 1\\
0 1 1& 0\\
1 0 0& 1\\
1 0 1& 1\\ 
1 1 0& 1\\
1 1 1& 1\\
\hline
\end{tabular} 
\begin{tabular}{|c | c|}
\hline
$x_1\ x_2\ x_3$&$f_3$\\\hline
0 0 0& 1\\
0 0 1& 0\\
0 1 0& 0\\
0 1 1& 1\\
1 0 0& 0\\
1 0 1& 0\\ 
1 1 0& 0\\
1 1 1& 0\\
\hline
\end{tabular} 
  \caption{Truth table for the Boolean functions in Example~\ref{ex:layers}.}\label{table:truthtable}
\end{table}
%
%
\subsection*{Definition of control actions}
This paper considers two types of control action: deletion of edges and constant expression of edges. An edge deletion represents the experimental intervention that prevents a regulation from happening. This action can be achieved by the use of therapeutic drugs that target a specific gene interaction, see reference~\cite{Choi2012} where this type of control has been experimentally applied.
A node deletion can be represented by the deletion of all of its outgoing edges. A constant expression or a constitutive activation of a node might result in aberrant cell proliferation and cancer, see~\cite{Choi2012} where the constant expression of \textit{cyclin G} in the signaling pathway of \textit{p53} is reported as a signature of abnormal gene expression that leads to cancer. But constant expressions could also help to drive the system into a more desirable state, see~\cite{Saadatpour:2011aa} where constant expression of nodes have been proposed as potential controls. As a proof of principle, this paper will consider the constant expression of an edge as a potential control action.
\begin{definition}[Edge Control]\label{def:edge_del}
Consider the edge $x_i\rightarrow x_j$ in the wiring diagram $\mathcal{W}$. For $u_{i,j}\in\mathbb F_2$, the control of the edge $x_i\rightarrow x_j$ consists of manipulating the input variable $x_i$ for $f_j$ in the following way: 
\begin{equation*}
\label{edge_del_def}
\mathcal{F}_j(x,u_{i,j}) = f_j(x_{j_1},\dots,(u_{i,j}+1)x_i,\dots,x_{j_m}).
\end{equation*}
For each value of $u_{i,j}$ we have the following control settings: 
\begin{itemize}
  \item When $u_{i,j}=0$, $\mathcal{F}_j(x,u_{i,j}) = f_j(x_{j_1},\dots,x_i,\dots,x_{j_m})$. That is, the control is not active.
  \item When $u_{i,j}=1$, $\mathcal{F}_j(x,u_{i,j}) = f_j(x_{j_1},\dots,x_i=0,\dots,x_{j_m})$. This is the case when the control is active and the action represents the removal of the edge $x_i\rightarrow x_j$.
\end{itemize}
\end{definition}
For simplicity, in Definition~\ref{def:edge_del} we considered only edge deletions. To include both the deletion and constant expression of an edge we could consider the following control function 
$$\mathcal{F}_j(x,\mu) = f_j(x_{j_1},\dots,(u^+_{i,j}+u^-_{i,j}+1)x_i+u^+_{i,j},\dots,x_{j_m})\text{ where }u^-_{i,j},u^+_{i,j}\in\mathbb F_2.$$
Then for each combination of $u^-_{i,j}$ and $u^+_{i,j}$ we have the following control settings: 
\begin{itemize}
  \item For $u^{-}_i=0, u^{+}_i=0$, $\mathcal{F}_j(x,0,0) = f_j(x_{j_1},\dots,x_i,\dots,x_{j_m})$. That is, the control is not active.
  \item For $u^{-}_i=1, u^{+}_i=0$, $\mathcal{F}_j(x,1,0)  = f_j(x_{j_1},\dots,x_i=0,\dots,x_{j_m})$. This action represents the knock out of the node $x_j$.
  \item For $u^{-}_i=0, u^{+}_i=1$, $\mathcal{F}_j(x,0,1)  = f_j(x_{j_1},\dots,x_i=1,\dots,x_{j_m})$. This action represents the constant expression of the node $x_j$.
  \item For $u^{-}_i=1, u^{+}_i=1$, $\mathcal{F}_j(x,1,1) =  f_j(x_{j_1},\dots,x_i+1,\dots,x_{j_m})$. This action changes the variable $x_i$ to its negative value and might not be a relevant case of control.
\end{itemize}
\section{Methods}
\subsection*{Eliminating state transitions through edge deletion and constant expression}
We avoid undesirable state transitions in the state space graph of a system of canalizing functions by means of edge deletion in the system's wiring diagram. 

Let $\mathbf{F}=(f_1,\ldots,f_n):\{0,1\}^n\to \{0,1\}^n$ be a Boolean network and $S=(V_s, E_s)$ be the state space graph of $\mathbf{F}$, where $V_s\subseteq \{0,1\}^n$ is the vertex set of $S$ and $E_s\subseteq \{0,1\}^n\times \{0,1\}^n$ is its edge set. Suppose for $\mathbf{u}, \mathbf{v}\in V_s$ there is a directed edge $\{\mathbf{u}, \mathbf{v}\} \in E_s$ which represents an undesirable transition. 
We eliminate the transition by deleting appropriate edges from the wiring diagram of the system, $W=(V_w, E_w)$, where $V_w=\{x_1,\ldots, x_n\}$ and $E_w=V_w\times V_w$. The following is a sufficient condition for eliminating a transition from $S$ through deleting an edge in $E_w$. 

\begin{theorem}\label{suff-cond}
Suppose $x_t \in V_w$ which takes input from $x_k \in V_w$, i.e. $\{x_k, x_t\} \in E_w$ (we will also use the notation $x_k \to x_t$). Let also $x_k$ be a canalizing variable in $f_t$, the functions that determines the state of $x_t$ in $S$. If the following four conditions are met, then deleting the edge $\{x_k, x_t\}$ from $E_w$ results in eliminating the transition $\{\mathbf{u}, \mathbf{v}\}$ from $E_s$:

\begin{enumerate}
\item No variable in a more dominant layer assumes its canalizing input in $\mathbf{u}$.
\item $x_k$ has canalizing input 0.
\item The $k$-th entry of $\mathbf{u}$ is 1, i.e. $[\mathbf{u}]_k=1$.
\item $x_k$ has canalizing output that is the negation of the $t$-th entry of $\mathbf{v}$, that is $\overline{[\mathbf{v}]_t}$.
\end{enumerate}
\end{theorem}

The reason behind the first condition is that if any variable whose layer is more or equally dominant than $x_k$'s layer has assumed its canalizing input in $\mathbf{u}$, then replacing $x_k$ with 0 will have no effect on $f_t$'s output. Deleting an edge has to impose change on the network for control and so the second requirement is needed since if $[\mathbf{u}]_k=0$ already, then deleting the edge $x_k\to x_t$ will have no effect on the network. The third condition has a similar explanation. 

Similar sufficient conditions can be stated for eliminating a state space transition through constant expression of an edge, simply by replacing 0 with 1 and 1 with 0.

Node deletion can also be used for control through canalization. In that case, node deletion corresponds to deleting the outgoing edges from the deleted node and Method~\ref{suff-cond} can be applied to each one of them.
\subsection*{Effect of edge deletion and constant expression on the state space}
We count the maximum number of state space transitions that can be changed as a result of deleting a single edge.

\begin{theorem}\label{bound}
Let $\mathbf{F}=(f_1,\ldots,f_n):\{0,1\}^n\to \{0,1\}^n$ be a Boolean network where $f_t$ is a PNCF of depth $d$ in $m$ variables in canalizing variable order $1, 2,\ldots, d$. The deletion of the edge $x_k\to x_t$ results in 

\begin{enumerate}
\item[(a)] up to $2^{n-\ell_1-\ell_2-...-\ell_{r}}$ changes in the state space if $k\le d$ and $x_k$ is in the $r$-th layer of $f_t$, where $\ell_1,\ldots \ell_r$ are the numbers of variables in layers $1,\ldots,r$, respectively; that is, the probability that any transition will be removed from the state space upon deletion of $x_k\to x_t$ is at most $2^{n-\ell_1-\ell_2-\ldots-\ell_{r}}/2^n=(\frac{1}{2})^{\ell_1+\ell_2+\ldots+\ell_{r}}$;
\item[(b)] up to $2^{n-d-1}$ changes in the state space if $d < k\le m$, i.e. $x_k$ is not canalizing; thus the probability that a particular transition will be removed from the state space upon deletion of $x_k\to x_t$ is at most $(\frac{1}{2})^{d-1}$.
\end{enumerate}
\end{theorem}

To see how the bound is calculated, notice that when $x_k\to x_t$ is deleted, half of the transition table can potentially change (the other half had $x_k=0$ already). Of the remaining half, half contains the canalizing input of a variable in the most dominant layer and so $x_k$ cannot cause change. Now half of the half only can possibly change but half of that has the canalizing input of another variable from a more or equally dominant layer to the one where $x_k$ is, thus preventing $x_k$ from causing change, etc.

This upper bound remains the same when instead of deleting an edge, an edge is constantly expressed.
\subsection*{General procedure for identifying control edges}
\begin{figure}[h!]
\begin{center}
\includegraphics[width=0.8\textwidth]{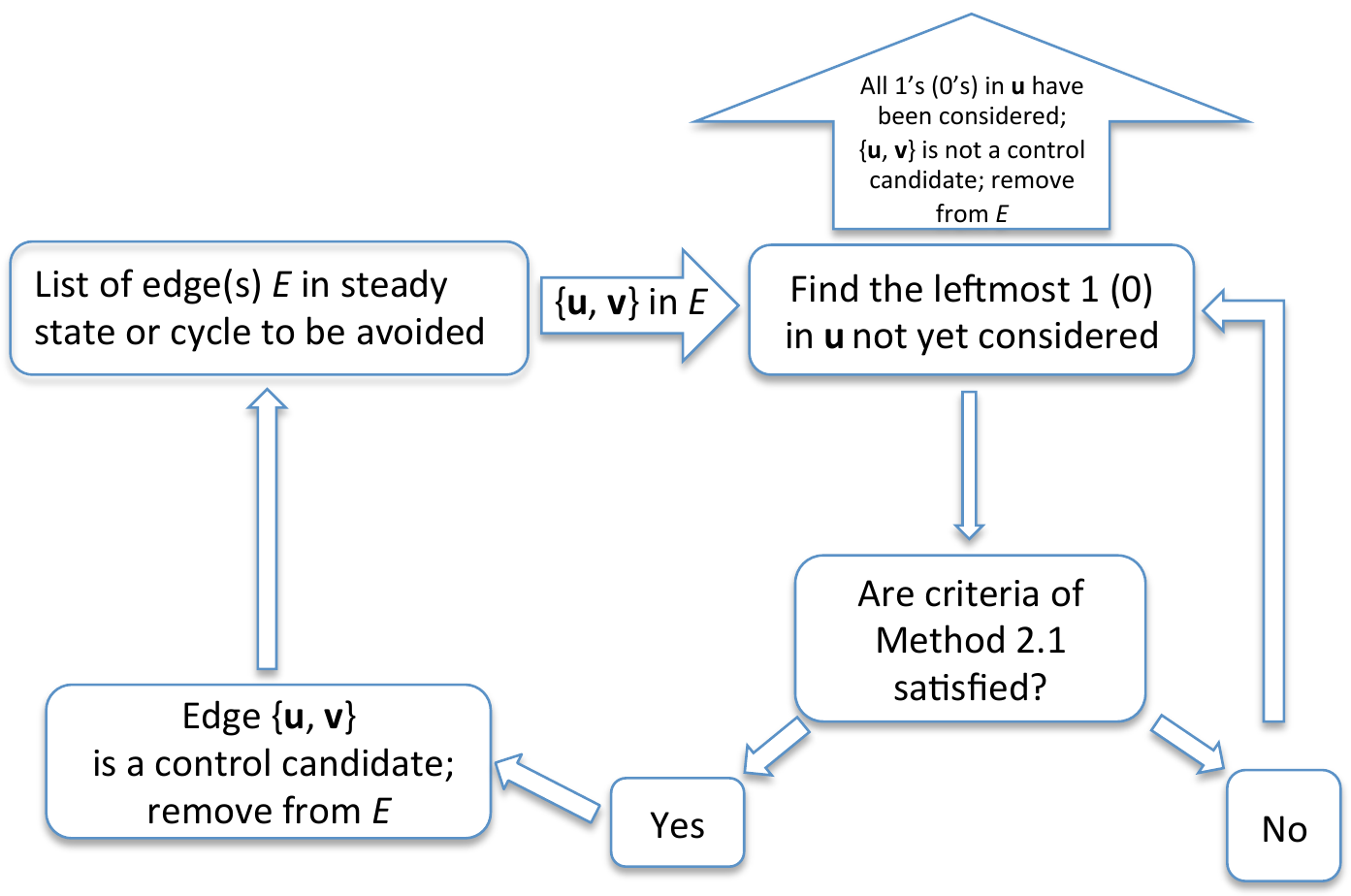}
\end{center}
  \caption{Flowchart of the procedure for identifying edges as control candidates.}
      \label{fc}
\end{figure}

Below we provide a general procedure for identifying control edges based on the two methods we developed. Figure~\ref{fc} further illustrates it.

Given a Boolean network model for a biological system:
\begin{enumerate}
\item Formulate a goal in terms of the part of the state space you wish to avoid, e.g. a fixed point or a cycle. If it contains more than one state, consider all states that are part of it. Choose to begin with one of them, $\mathbf{v}$.
\item For control via edge deletion, identify all 1's in the state $\mathbf{u}$ preceding $\mathbf{v}$; for control via constant expression, identify all 0's.
\item Begin with, say, the leftmost 1 (or 0) in $\mathbf{u}$. Observe the canalizing structure of the functions in your model.
Then check if the conditions of Method~\ref{suff-cond} are satisfied.
\item If some of the conditions of Method~\ref{suff-cond} are not met or you are looking for other control options, proceed to the next 1 (or 0) in $\mathbf{u}$.
\item If you wish to avoid a cycle or other trajectory that contains several states, you can repeat the above steps on all states in order to find all control options.
\item If you find multiple edges as candidates for control, you may want to choose to delete the ones that have the smallest impact on the state space thus minimizing the side effects of edge manipulation.
\end{enumerate}

\section*{Results}

We apply the control methods we developed to the Boolean models of two networks: a model of the human tumor suppressor gene \textit{p53} pathways \cite{layek-2011} and a mammalian cell-cycle network~\cite{Faure:2006kq}.

\subsection*{\textit{p53-mdm2} model}
In \cite{layek-2011}, a Boolean model, Eq. (\ref{p53-eqns}), of the widely studied \textit{p53} pathway is built, where the external signal is dna\_dsb, the DNA damage input. 

\begin{eqnarray}\label{p53-eqns}
\begin{aligned}
\rm{ATM}_{next}&=\overline{\rm{Wip1}}(\rm{ATM}+\rm{dna\_dsb})\\
\rm{p53}_{next}&=\overline{\rm{Mdm2}}(\rm{ATM}+\rm{Wip1})\\
\rm{Wip1}_{next}&=\rm{p}53\\
\rm{Mdm2}_{next}&=\overline{\rm{ATM}}(\rm{p53}+\rm{Wip1})
\end{aligned}
\end{eqnarray}

\begin{figure}[h!]
\begin{center}
\includegraphics[width=0.4\textwidth]{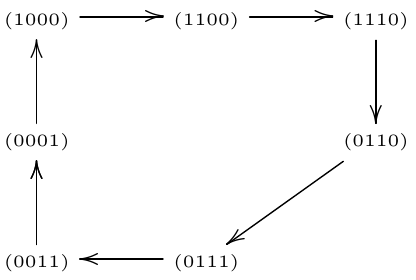}
\end{center}
  \caption{The cycle for the Boolean network of the \textit{p53} pathways in the presence of DNA damage. The state space is defined as (ATM p53 Wip1 Mdm2).}
      \label{p53-cycle}
      \end{figure}

The other variables are ATM, p53, Wip1, and Mdm2. When dna\_dsb $=0$, the state space has a single fixed point, $(0 0 0 0)$, corresponding to no stress. However, when dna\_dsb $=1$, i.e. the DNA damage input turns on, the state space contains a single cycle of length seven (Fig.~\ref{p53-cycle}) and no fixed points. The cycle represents cyclic variation in the expression patterns of all the four genes.
We want to prevent this cycle from taking place through removing one or multiple transitions from it.

\begin{figure}[h!]
\begin{center}
\includegraphics[width=0.4\textwidth]{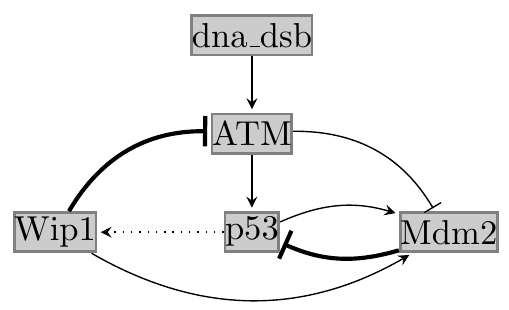}
\end{center}
  \caption{Wiring diagram of the \textit{p53-mdm2} model from \cite{layek-2011}. Pointed arrows correspond to positive regulatory interactions and blunt arrows to negative regulatory interactions. The edge deletion control candidate is given as a dotted edge and the two constant expression control candidates are denoted by thicker edges.}
      \label{p53-wd}
\end{figure}

%
%
The wiring diagram of the model is presented on Fig.~\ref{p53-wd}. By Method~\ref{suff-cond}, we identify that deleting edge p53$\to$Wip1 (which also happens to correspond to deleting the node Wip1) has the effect of removing the following four transitions from the undesirable cycle in Fig.~\ref{p53-cycle}: 

\begin{eqnarray*}
(1 \mathbf{1} 0 0) \to (1 1 \mathbf{1} 0), \\
(1 \mathbf{1} 1 0) \to (0 1 \mathbf{1} 0), \\
(0 \mathbf{1} 1 0) \to (0 1 \mathbf{1} 1), \\
(0 \mathbf{1} 1 1) \to (0 0 \mathbf{1} 1).
\end{eqnarray*}

The bold entries in the states correspond to the entries where conditions 3 and 4 of Method~\ref{suff-cond} apply. As a result, the system has only a single steady state, $(1 1 0 0 )$. Unfortunately, not all four proteins are inactive in it but we will see that this can be achieved through constant expression of the edge Wip1$\to$ATM, as it will be discussed later. 

We can also count the number of changes that deleting p53$\to$Wip1 from the wiring diagram induces on the state space transitions: by Method~\ref{bound}, there can be up to $2^{4-1}=8$ changes and in fact we observe exactly as many, demonstrating that the bound from Method~\ref{bound} is sharp: out of the 16 states in the state space, eight contain p53 $=1$. When applying the original update rules, the value of p53 in these eight states remains 1 (and so Wip1 becomes 1), while after deleting p53$\to$Wip1, Wip1 becomes 0 since now the output of its update rule is 0, thus causing a change in the state space.

Constant expression of an edge is another strategy for removal of transitions in a state space graph. There are analogous conditions to Method~\ref{suff-cond} for constant expression, obtained by simply replacing 0 with 1 and vice versa. For example, Wip1 is a canalizing variable in the function of ATM with canalizing input 1 and canalizing output 0. Therefore, we can set the edge Wip1$\to$ATM to constant expression in order to remove the following transitions from the undesirable cycle: $(1 0 0 0 1) \to (1 1 0 0 1), (1 1 0 0 1) \to (1 1 1 0 1)$, and  $(0 0 0 1 1)\to (1 0 0 0 1)$. The result is a state space with fixed point $(0 0 0 0)$, corresponding to no stress as when dna\_dsb $=0$. Another option for control via constant expression is the edge Mdm2$\to$p53 which also results in a single steady state, although this time it is $(1 0 0 0)$.

\subsection*{Mutated cell-cycle network model}
\begin{figure}[h!]
\begin{center}
\includegraphics[width=0.4\textwidth]{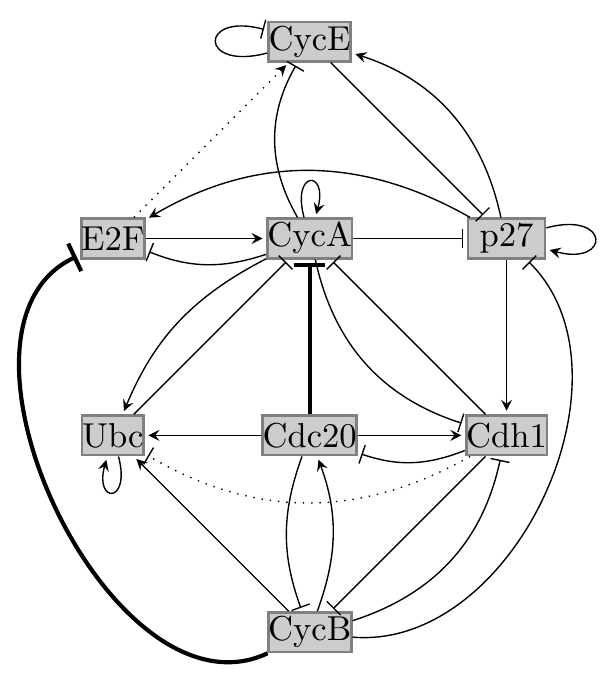}
\end{center}
  \caption{Wiring diagram of the mutated (CycD and Rb inactive) mammalian cell-cycle network. Pointed arrows correspond to positive regulatory interactions and blunt arrows to negative regulatory interactions~\cite{Faure:2006kq}. Edge deletion control candidates are dotted and constant expression control candidates are denoted by thicker edges.}
      \label{cell-cycle-wd}
      \end{figure}

\begin{figure}[h!]
\begin{center}
\includegraphics[width=0.4\textwidth]{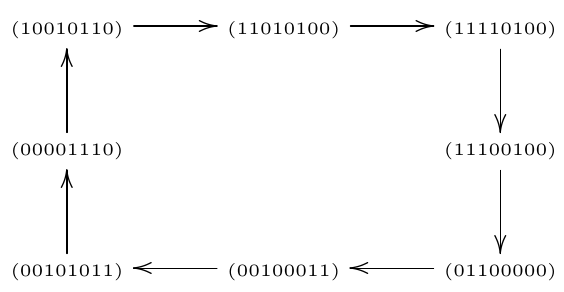}
\end{center}
  \caption{The cycle which persists in the absence of a growth factor in the mutated phenotype (Rb inactive). The state space is defined as (E2F  CycE CycA p27 Cdc20 Cdh1 Ubc CycB).}
      \label{bad-cycle}
      \end{figure}

As a second application, we consider Faur\'{e} et al.~\cite{Faure:2006kq} who proposed a Boolean model of the cell cycle progression. We focus on the scenario when the tumor suppressor retinoblastoma protein Rb is absent as reflected in Eq. (\ref{faure-eqns}). The wiring diagram for that case is given in Fig. \ref{cell-cycle-wd}. Faur\'{e} et al.~\cite{Faure:2006kq} assume that the expression of CycD changes independently of the cell's content and reflects the state of the growth factor. According to their model, the mammalian cell cycle with a mutated phenotype will cycle through the eight states (Fig. \ref{bad-cycle}) even when CycD is inactive. 

\begin{eqnarray}\label{faure-eqns}
\begin{aligned}
\rm{E2F} &= (\overline{\rm{CycA}} \land \overline{\rm{CycB}}) \lor (\rm{p27} \land \overline{\rm{CycB}})\\
\rm{CycE} &= \rm{E2F}\\
\rm{CycA} &= \rm{(E2F \land \overline{Cdc20} \land \overline{(Cdh1 \land Ubc)})} \\
& ~~~~~\lor \rm{(CycA \land \overline{Cdc20} \land \overline{(Cdh1 \land Ubc)})}\\
\rm{p27} &= \rm{(p27 \land \overline{(CycE \land CycA)} \land \overline{CycB} \land \overline{CycD}})\\
& ~~~~~ \lor \rm{(\overline{CycE} \land \overline{CycA} \land \overline{CycB}) }\\
\rm{Cdc20} &= \rm{CycB}\\
\rm{Cdh1} &= \rm{(\overline{CycA} \land \overline{CycB}) \lor Cdc20 \lor (p27 \land \overline{CycB})}\\
\rm{Ubc} &= \rm{\overline{Cdh1} \lor (Cdh1 \land Ubc} \\
& ~~~~~\rm{\land (Cdc20 \lor CycA \lor CycB)) }\\
\rm{CycB} &= \rm{\overline{Cdc20} \land \overline{Cdh1}}
\end{aligned}
\end{eqnarray}

We propose four edges from the wiring diagram in Fig.~\ref{cell-cycle-wd} that can be used for control in order to avoid the cycle in Fig.~\ref{bad-cycle}. These edges were identified following Method~\ref{suff-cond} applied on transitions in the cycle with the objective of eliminating the cycle and also leading the system to fixed point(s) where $\rm{p27=Cdh1=1}$ as in a normal cell. The results are summarized in Table~\ref{cycle-results}. Other attempts at control produced fixed points and/or cycles where $\rm{p27=Cdh1=0}$.
\begin{table}[!htb]
\centering
\begin{tabular}{|l|l|}
\hline
\small{Control action} & \small{Outcome} \\
\hline
\small{Delete E2F$\to$CycE} & \small{Single fixed point} \\ & \small{with $\rm{p27=Cdh1=1}$}\\
& \\
\small{Delete Cdh1$\to$Ubc} & \small{$>90\%$ of states in component} \\ & \small{with fixed point where} \\& \small{$\rm{p27=Cdh1=1}$}\\
& \\
\small{Express constantly} & \small{Two fixed points}, \\ \small{CycB$\to$E2F} & \small{both with $\rm{p27=Cdh1=1}$}\\
& \\
\small{Express constantly} & \small{$> 90\%$ of states in component} \\ \small{Cdc20$\to$CycA} & \small{with fixed point where} \\& \small{$\rm{p27=Cdh1=1}$}\\
\hline
\end{tabular}
\caption{Summary of control results for the mutated cell-cycle network.}
\label{cycle-results}
\end{table}

\section*{Discussion}
In the Methods section, we noticed that the bound from Method~\ref{bound} is sharp. This was demonstrated using the \textit{p53} model. In general, the exact bound from part \emph{(a)} of Method~\ref{bound} is achieved when $\mathcal{F}_t(x,u_{t,k})=x_k$ or $\overline{x_k}$ (as a function), where $\mathcal{F}_t(x,u_{t,k})$ is the function obtained from $f_t$ by plugging in the canalizing input of the variables that are more or equally dominant to $x_k$.

The bound from Method~\ref{bound} can also help choose which edge to delete or constantly express when there is more than one option with the purpose of controlling the side effects resulting from an edge manipulation. If it is desirable to minimize the impact on the state space, thus avoiding possible negative side effects on the system, one should choose for control an edge whose input variable is in the least dominant layer possible in the target function. That is, if $x_i$ and $x_j$ are both canalizing variables in $f_t$ and $x_j$ is in a less dominant layer than $x_i$, then one should choose to delete or constantly express the edge $x_j \to x_t$ since according to the bound of Method~\ref{bound}, the maximum impact of this control is smaller than if $x_i \to x_t$ is manipulated.

It is important to point out that Method~\ref{suff-cond} only guarantees that a certain transition will be avoided and one may be able to use this to remove a cycle from a state space as we did for the \textit{p53} model. However, the method does not guarantee that the system will not contain other cycles since removing transitions from a cycle destroys the cycle but may also create a different one; nor does it guarantee that the resulting fixed point will be exactly the desired one as it was observed in both applications. To find controllers that give the desired fixed points one could use the algebraic methods described in~\cite{algcontrol2015}.

This paper considers edge manipulations as potential control actions to avoid undesirable attractors. Control through edge manipulations in the wiring diagram of a network has been previously considered in~\cite{Campbell:2014aa}. Although the authors of~\cite{Campbell:2014aa} consider edge additions in the wiring diagram as control actions, that is, by adding new regulators to the existing set of regulators that help the system to transition into a desirable attractor.

\section*{Conclusions}

The structure of the canalizing variables in a biologically relevant Boolean rule plays an important role in the control of Boolean networks. 
Special combinations of canalizing inputs can help identify network controllers and the canalizing structure of a Boolean function allows to estimate the number of transitions that change after using the type of controllers proposed in this paper. Moreover, the hierarchy of the canalizing variables can be used for assessing the impact on the network dynamics as a result of a given control.
This paper exploits the canalizing properties of Boolean rules to derive a method that can be useful for identifying control targets for avoiding undesirable states.
Additionally, it provides a method for assessing the impact of the controllers on the dynamics of the uncontrolled network.
Thus these two complementary methods can help in the selection of appropriate controllers. Method~\ref{suff-cond} gives a practical way for identifying the potential edges to control in the wiring diagram of a network for avoiding undesirable state transitions. Method~\ref{bound}, on the other hand, provides a measure of the impact of an edge deletion onto the state space of a model and establishes that this impact differs significantly based on the canalizing properties of the nodes involved: an edge coming from a node with stronger canalization, represented in the model by a variable in a more dominant layer, has exponentially higher probability to change the state space than an edge from a node with weaker canalization, represented by a variable in a less dominant layer. Therefore, Method~\ref{bound} is a useful tool for assessing the impact of the controllers identified by Method~\ref{suff-cond} on the dynamics of the system providing a way for selecting desirable controllers.
\section*{Competing interests}
The authors declare that they have no competing interests.

\section*{Author's contributions}
The two authors contributed equally to all aspects of the paper.

\section*{Acknowledgements}
The second author was supported by NSF Award \#1419038.
\bibliographystyle{siam}
\bibliography{article_david}      

\begin{thebibliography}{10}

\bibitem{Abou-Jaoude:2009aa}
{\sc W.~Abou-Jaoud{\'e}, D.~A. Ouattara, and M.~Kaufman}, {\em From structure
  to dynamics: frequency tuning in the p53-mdm2 network i. logical approach}, J
  Theor Biol, 258 (2009), pp.~561--77.

\bibitem{Albert:2003aa}
{\sc R.~Albert and H.~G. Othmer}, {\em The topology of the regulatory
  interactions predicts the expression pattern of the segment polarity genes in
  drosophila melanogaster}, J Theor Biol, 223 (2003), pp.~1--18.

\bibitem{Balleza:2008aa}
{\sc E.~Balleza, E.~R. Alvarez-Buylla, A.~Chaos, S.~Kauffman, I.~Shmulevich,
  and M.~Aldana}, {\em Critical dynamics in genetic regulatory networks:
  examples from four kingdoms}, PLoS One, 3 (2008), p.~e2456.

\bibitem{Campbell:2014aa}
{\sc C.~Campbell and R.~Albert}, {\em Stabilization of perturbed boolean
  network attractors through compensatory interactions}, BMC Syst Biol, 8
  (2014), p.~53.

\bibitem{Choi2012}
{\sc M.~Choi, J.~Shi, S.~H. Jung, X.~Chen, and K.-H. Cho}, {\em Attractor
  landscape analysis reveals feedback loops in the p53 network that control the
  cellular response to dna damage}, Sci. Signal., 5 (2012), p.~ra83.

\bibitem{Davidich:2008aa}
{\sc M.~I. Davidich and S.~Bornholdt}, {\em Boolean network model predicts cell
  cycle sequence of fission yeast}, PLoS One, 3 (2008), p.~e1672.

\bibitem{Erler:2012zr}
{\sc J.~T. Erler and R.~Linding}, {\em Network medicine strikes a blow against
  breast cancer}, Cell, 149 (2012), pp.~731--3.

\bibitem{Faure:2006kq}
{\sc A.~Faur{\'e}, A.~Naldi, C.~Chaouiya, and D.~Thieffry}, {\em Dynamical
  analysis of a generic boolean model for the control of the mammalian cell
  cycle}, Bioinformatics, 22 (2006), pp.~e124--31.

\bibitem{He:2015aa}
{\sc Q.~He and M.~Macauley}, {\em Stratification and enumeration of boolean
  functions by canalizing depth},  (2015).

\bibitem{Helikar:2013aa}
{\sc T.~Helikar, N.~Kochi, B.~Kowal, M.~Dimri, M.~Naramura, S.~M. Raja,
  V.~Band, H.~Band, and J.~A. Rogers}, {\em A comprehensive, multi-scale
  dynamical model of erbb receptor signal transduction in human mammary
  epithelial cells}, PLoS One, 8 (2013), p.~e61757.

\bibitem{Helikar:2008aa}
{\sc T.~Helikar, J.~Konvalina, J.~Heidel, and J.~A. Rogers}, {\em Emergent
  decision-making in biological signal transduction networks}, Proc Natl Acad
  Sci U S A, 105 (2008), pp.~1913--8.

\bibitem{Huang:1999aa}
{\sc S.~Huang}, {\em Gene expression profiling, genetic networks, and cellular
  states: an integrating concept for tumorigenesis and drug discovery}, J Mol
  Med (Berl), 77 (1999), pp.~469--80.

\bibitem{Jarrah:2007aa}
{\sc A.~S. Jarrah, B.~Raposa, and R.~Laubenbacher}, {\em Nested canalyzing,
  unate cascade, and polynomial functions}, Physica D, 233 (2007),
  pp.~167--174.

\bibitem{Kauffman2003}
{\sc S.~Kauffman, C.~Peterson, B.~Samuelsson, and C.~Troein}, {\em Random
  boolean network models and the yeast transcriptional network}, Proceedings of
  the National Academy of Sciences, 100 (2003), pp.~14796--14799.

\bibitem{Kauffman:2004aa}
\leavevmode\vrule height 2pt depth -1.6pt width 23pt, {\em Genetic networks
  with canalyzing boolean rules are always stable}, Proc Natl Acad Sci U S A,
  101 (2004), pp.~17102--7.

\bibitem{Kauffman:1969aa}
{\sc S.~A. Kauffman}, {\em Metabolic stability and epigenesis in randomly
  constructed genetic nets}, J Theor Biol, 22 (1969), pp.~437--67.

\bibitem{layek-2011}
{\sc R.~K. Layek, A.~Datta, and E.~R. Dougherty}, {\em From biological pathways
  to regulatory networks}, Molecular bioSystems, 7 (2011), pp.~843---851.

\bibitem{Layne:2012aa}
{\sc L.~Layne, E.~Dimitrova, and M.~Macauley}, {\em Nested canalyzing depth and
  network stability}, 74 (2012), pp.~422--433.

\bibitem{Lee:2012aa}
{\sc M.~J. Lee, A.~S. Ye, A.~K. Gardino, A.~M. Heijink, P.~K. Sorger,
  G.~MacBeath, and M.~B. Yaffe}, {\em Sequential application of anticancer
  drugs enhances cell death by rewiring apoptotic signaling networks}, Cell,
  149 (2012), pp.~780--94.

\bibitem{Li:2004aa}
{\sc F.~Li, T.~Long, Y.~Lu, Q.~Ouyang, and C.~Tang}, {\em The yeast cell-cycle
  network is robustly designed}, Proc Natl Acad Sci U S A, 101 (2004),
  pp.~4781--6.

\bibitem{Li:2013aa}
{\sc Y.~Li, J.~O. Adeyeye, D.~Murrugarra, B.~Aguilar, and R.~Laubenbacher},
  {\em Boolean nested canalizing functions: A comprehensive analysis},
  Theoretical Computer Science, 481 (2013), pp.~24--36.

\bibitem{Mendoza:2006aa}
{\sc L.~Mendoza}, {\em A network model for the control of the differentiation
  process in th cells}, Biosystems, 84 (2006), pp.~101--14.

\bibitem{Murrugarra:2011aa}
{\sc D.~Murrugarra and R.~Laubenbacher}, {\em Regulatory patterns in molecular
  interaction networks}, Journal of Theoretical Biology, 288 (2011),
  pp.~66--72.

\bibitem{Murrugarra:2012aa}
\leavevmode\vrule height 2pt depth -1.6pt width 23pt, {\em The number of
  multistate nested canalyzing functions}, Physica D: Nonlinear Phenomena, 241
  (2012), pp.~929--938.

\bibitem{Murrugarra2012}
{\sc D.~Murrugarra, A.~Veliz-Cuba, B.~Aguilar, S.~Arat, and R.~Laubenbacher},
  {\em Modeling stochasticity and variability in gene regulatory networks},
  EURASIP Journal on Bioinformatics and Systems Biology, 2012 (2012), p.~5.

\bibitem{algcontrol2015}
{\sc D.~Murrugarra, A.~Veliz-Cuba, B.~Aguilar, and R.~Laubenbacher}, {\em
  Identification of control targets of boolean molecular network models via
  computational algebra}, To appear,  (2015).

\bibitem{Saadatpour:2011aa}
{\sc A.~Saadatpour, R.-S. Wang, A.~Liao, X.~Liu, T.~P. Loughran, I.~Albert, and
  R.~Albert}, {\em Dynamical and structural analysis of a t cell survival
  network identifies novel candidate therapeutic targets for large granular
  lymphocyte leukemia}, PLoS Comput Biol, 7 (2011), p.~e1002267.

\bibitem{DBLP:books/daglib/0024105}
{\sc I.~Shmulevich and E.~R. Dougherty}, {\em Probabilistic Boolean Networks -
  The Modeling and Control of Gene Regulatory Networks}, {SIAM}, 2010.

\bibitem{Thomas:1990aa}
{\sc R.~Thomas and R.~D'Ari}, {\em Biological feedback}, CRC Press, Boca Raton,
  1990.

\bibitem{DBLP:journals/jcb/Veliz-CubaS11}
{\sc A.~Veliz{-}Cuba and B.~Stigler}, {\em Boolean models can explain
  bistability in the \emph{lac} operon}, Journal of Computational Biology, 18
  (2011), pp.~783--794.

\bibitem{waddington1942}
{\sc C.~H. Waddington}, {\em Canalization of development and the inheritance of
  acquired characters}, Nature, 150 (1942), pp.~563--565.

\bibitem{Wang:2013aa}
{\sc W.~Wang}, {\em Therapeutic hints from analyzing the attractor landscape of
  the p53 regulatory circuit}, Sci Signal, 6 (2013), p.~pe5.

\bibitem{Wang:2014qy}
{\sc Z.~Wang and T.~S. Deisboeck}, {\em Mathematical modeling in cancer drug
  discovery}, Drug Discovery Today, 19 (2014), pp.~145--150.

\bibitem{Zhang:2008aa}
{\sc R.~Zhang, M.~V. Shah, J.~Yang, S.~B. Nyland, X.~Liu, J.~K. Yun, R.~Albert,
  and T.~P. Loughran, Jr}, {\em Network model of survival signaling in large
  granular lymphocyte leukemia}, Proc Natl Acad Sci U S A, 105 (2008),
  pp.~16308--13.

\end{thebibliography}
\end{document}